\documentclass{article}
\usepackage{cite} 
\usepackage{amsmath} 
\usepackage{mathtools} 
\usepackage{bm} 
\usepackage{float}
\usepackage[dvipsnames]{xcolor}
\usepackage{amsfonts}
\usepackage{graphicx}
\usepackage{epstopdf}
\usepackage{newfloat}
\usepackage{tocloft}
\tocloftpagestyle{empty}
\DeclareFloatingEnvironment[name={Supporting Information Figure},fileext=lsf, listname={List of Supporting Information Figures}]{suppfigure}

\usepackage{authblk}
\usepackage[font=small]{caption}

\usepackage{booktabs}
\usepackage{multirow}       
\usepackage{tabularx}       
\usepackage{hhline}
\usepackage{arydshln}		
\usepackage{xcolor}
\let\citeleft=(
\let\citeright=)

\newtagform{brackets}{[}{]}
\usetagform{brackets}

\usepackage[colorinlistoftodos,prependcaption,textsize=small]{todonotes} 
\presetkeys%
{todonotes}%
{noline, backgroundcolor=yellow}{}
\tikzset{/tikz/notestyleraw/.append style={text=blue, font=\bfseries}}
\long\def\fb#1{\todo[noline, disable]{#1}}
\long\def\fbb#1{\todo[noline, disable]{#1}}

\begin{document}

\pdfinfo{
	/Author (AUTHORS)
	/Title (TITLE)
}

\makeatletter 
\renewcommand\@biblabel[1]{#1.} 
\makeatother

\definecolor{xiaojian}{rgb}{0, 0, 0}
\definecolor{yablonskiy}{rgb}{0, 0, 0}
\definecolor{todo}{rgb}{0, 0, 0}

\definecolor{xiaojian2}{rgb}{0, 0, 0}

\def\ebf{{\mathbf{e}}}
\def\rbf{{\mathbf{r}}}
\def\sbf{{\mathbf{s}}}
\def\xbf{{\mathbf{x}}}
\def\ybf{{\mathbf{y}}}
\def\fbf{{\mathbf{f}}}
\def\pbf{{\mathbf{p}}}
\def\ubf{{\mathbf{u}}}

\def\rbfhat{{\widehat{\mathbf{r}}}}
\def\sbfhat{{\widehat{\mathbf{s}}}}
\def\pbfhat{{\widehat{\mathbf{p}}}}
\def\xbfhat{{\widehat{\mathbf{x}}}}
\def\Rbfhat{{\widehat{\mathbf{R}}}}
\def\Sbfhat{{\widehat{\mathbf{S}}}}

\def\Ical{{\mathcal{I}}}
\def\Dcal{{\mathcal{D}}}
\def\Lcal{{\mathcal{L}}}
\def\Mcal{{\mathcal{M}}}
\def\Bcal{{\mathcal{B}}}

\def\rbftilde{{\tilde{\mathbf{r}}}}
\def\sbftilde{{\tilde{\mathbf{s}}}}
\def\pbftilde{{\tilde{\mathbf{p}}}}
\def\ybftilde{{\tilde{\mathbf{y}}}}
\def\ubftilde{{\tilde{\mathbf{u}}}}

\def\thetabm{{\bm{\theta}}}
\def\etabm{{\bm{\eta}}}
\def\nubm{{\bm{\nu}}}

\def\Ibf{{\mathbf{I}}}
\def\Sbf{{\mathbf{S}}}
\def\Fbf{{\mathbf{F}}}
\def\Tbf{{\mathbf{T}}}
\def\Bbf{{\mathbf{B}}}
\def\Rbf{{\mathbf{R}}}
\def\Mbf{{\mathbf{M}}}

\definecolor{lightgreen}{rgb}{.9,1,.9}
\newcolumntype{L}[1]{>{\raggedright\arraybackslash}p{#1}}
\newcolumntype{C}[1]{>{\centering\arraybackslash}p{#1}}
\newcolumntype{R}[1]{>{\raggedleft\arraybackslash}p{#1}}
\newcolumntype{M}[1]{>{\centering\arraybackslash}m{#1}}

\newcommand{\rmk}[1]{%
	{\textcolor{red}{[\small #1]}}%
}

\newcommand{\eq}[1]{Eq.~(\ref{eq:#1})}
\newcommand{\fig}[1]{Fig.~\ref{fig:#1}}
\newcommand{\figs}[1]{Figs.~\ref{fig:#1}}
\newcommand{\fign}[1]{\ref{fig:#1}}
\newcommand{\tbl}[1]{\ref{tab:#1}}
\newcommand{\sctn}[1]{Sec.~\ref{sec:#1}}

\newcommand{\mb}[1]{\mathbf{#1}}
\newcommand{\mbs}[1]{\boldsymbol{#1}}
\newcommand{\mbb}[1]{\mathbb{#1}}
\newcommand{\mc}[1]{\mathcal{#1}}

\providecommand{\abs}[1]{\left\lvert#1\right\rvert}
\providecommand{\norm}[1]{\left\lVert#1\right\rVert}
\DeclarePairedDelimiterX{\normsz}[1]{\lVert}{\rVert}{#1}

\title{\vspace{-2cm} Learning-based Motion Artifact Removal Networks (LEARN) for Quantitative $R_2^\ast$ Mapping} 

\author[1]{Xiaojian~Xu}
\author[2]{ Satya~V.~V.~N.~Kothapalli}
\author[3]{Jiaming~Liu}
\author[2]{Sayan~Kahali}
\author[1]{Weijie~Gan}
\author[2]{Dmitriy~A.~Yablonskiy}
\author[1,3]{Ulugbek~S.~Kamilov}

\affil[1]{\small Department of Computer Science and Engineering, Washington University in St.~Louis, St.~Louis, MO 63130, USA}
\affil[2]{\small Department of Radiology, Washington University in St.~Louis, St.~Louis, MO 63110, USA}
\affil[3]{\small Department of Electrical and Systems Engineering, Washington University in St.~Louis, St.~Louis, MO 63130, USA}
\maketitle

\vfill
\noindent
\textit{Running head:} Learning-based Motion Artifact Removal Networks (LEARN) for Quantitative $R_2^\ast$ Mapping

\noindent
\textit{Address correspondence to:} \\
Ulugbek~S.~Kamilov, One Brookings Drive, MSC 1045-213-1010J, St.~Louis, MO 63130, USA.\\
Email: kamilov@wustl.com

\noindent
This work was supported in part by the NSF CAREER award CCF-2043134, NIH/NIA grant  R01AG054513, Marilyn Hilton Award for Innovation in MS Research, and NVIDIA Corporation with the donation of the Titan Xp GPU for research.

\noindent
Approximate word count: 250 (abstract)  5,293 (body)\\

\noindent
Submitted to \emph{Magnetic Resonance in Medicine} as a Research Article.\\

\clearpage

\section*{Abstract}

\noindent
\textbf{Purpose}: To introduce two novel learning-based motion artifact removal networks (LEARN) for the estimation of quantitative motion- and $B0$-inhomogeneity-corrected $R_2^\ast$ maps from motion-corrupted multi-Gradient-Recalled Echo (mGRE) MRI data. 

\noindent
\textbf{Methods}: We \emph{train} two convolutional neural networks (CNNs) to correct motion artifacts for high-quality estimation of quantitative $B0$-inhomogeneity-corrected $R_2^\ast$ maps from mGRE sequences. The first CNN, LEARN-IMG, performs motion correction on complex mGRE images, to enable the subsequent computation of high-quality motion-free quantitative $R_2^\ast$ (and any other mGRE-enabled) maps using the standard voxel-wise analysis or machine-learning-based analysis. The second CNN, LEARN-BIO, is trained to directly generate motion- and $B0$-inhomogeneity-corrected quantitative $R_2^\ast$ maps from motion-corrupted magnitude-only mGRE images by taking advantage of the biophysical model describing the mGRE signal decay.

\noindent
\textbf{Results}: We show that both CNNs trained on synthetic MR images are capable of suppressing motion artifacts while preserving details in the predicted quantitative $R_2^\ast$ maps. Significant reduction of motion artifacts on experimental in vivo motion-corrupted data has also been achieved by using our trained models. 

\noindent
\textbf{Conclusion}: Both LEARN-IMG and LEARN-BIO can enable the computation of high-quality motion- and $B0$-inhomogeneity-corrected $R_2^\ast$ maps. LEARN-IMG performs motion correction on mGRE images and relies on the subsequent analysis for the estimation of $R_2^\ast$ maps, while LEARN-BIO directly performs motion- and $B0$-inhomogeneity-corrected $R_2^\ast$ estimation. Both LEARN-IMG and LEARN-BIO jointly process all the available gradient echoes, which enables them to exploit spatial patterns available in the data. The high computational speed of LEARN-BIO is an advantage that can lead to a broader clinical application.

\noindent
\textbf{Keywords}: MRI, motion correction,  convolutional neural networks, deep learning, gradient recalled echo, $R_2^\ast$ mapping, self-supervised deep learning.

\clearpage

\section*{Introduction}
\label{sec:introduction}
Multi-Gradient-Recalled-Echo (mGRE) sequences accompanied by correction of magnetic field inhomogeneity artifacts~\cite{Yablonskiy.etal2013, Hernando.etal2012} are used in different MRI applications to produce quantitative maps related to biological tissue microstructure in health and disease (e.g. \cite{ Hernando.etal2012, Zhao.etal2016,Ulrich.Yablonskiy2016, Zhao.etal2017a, Wen.etal2018a,  xiang2020quantitative, Kothapalli.etal2021, Roberts.etal2021}). However, involuntary physical motion and subtle anatomical fluctuations during the mGRE signal acquisition can lead to undesirable artifacts during the estimation of these quantitative maps. It is therefore important to develop methods that reduce the sensitivity of the estimated quantitative maps to the motion artifacts in the MR images. 

A number of methods have been developed over the years for the prevention, mitigation, or correction of motion artifacts in MR images~\cite{Odille.etal2008, Ooi.etal2009, White.etal2010,  Lingala.etal2011, Tisdall.etal2012,  Maclaren.etal2012, Loktyushin.etal2013, Wen.etal2015, Roujol.etal2015, Gallichan.etal2016, Cordero-Grande.etal2018, Haskell.etal2018, Johnson.Drangova2019}. Deep learning (DL) methods have  also been recently introduced for motion-correction in MRI due to their speed and quality of reconstruction~\cite{Haskell.etal2019, Kustner.etal2019, Armanious.etal2020, Sommer.etal2020, Liu.etal2020a, Oh.etal2021}. Despite the recent activity, DL is yet to be investigated in the context of quantitative $B0$-inhomogeneity-corrected estimation of $R_2^\ast$ maps from mGRE signals. One of the  key challenges in this context is the sensitivity of the quantitative maps to the motion artifacts~\cite{Callaghan.etal2015}.

In this paper, we propose two convolutional neural networks (CNNs) for recovering high-quality quantitative $R_2^\ast$ maps from the motion-corrupted mGRE images. Both of our methods, referred to LEARN-IMG and LEARN-BIO, are trained on motion-free mGRE images and their simulated motion-corrupted counterparts. LEARN-IMG follows the traditional supervised training strategy in order to correct the motion on the complex mGRE images. The high-quality motion-free and $B0$-inhomogeneity-corrected $R_2^\ast$  maps can be subsequently computed by applying the standard non-linear least squares (NLLS) analysis that also accounts for the effect of background $B0$ field gradients (herein we use Voxel Spread Function (VSF) approach~\cite{Yablonskiy.etal2013}) on the motion-corrected output images. On the other hand, LEARN-BIO is trained to directly map the magnitude-only motion-corrupted mGRE images to motion-free and and $B0$-inhomogeneity-corrected $R_2^\ast$ maps. The key feature of LEARN-BIO is that it is fully self-supervised, in the sense that it does not need ground-truth quantitative $R_2^\ast$ maps for training. Instead, it is trained using only the mGRE images and the biophysical model connecting the mGRE signal with biological tissue microstructure that includes contribution of magnetic field inhomogeneities to the mGRE signal decay (described in terms of a factor $F(t)$~\cite{Yablonskiy.etal2013}), and our knowledge of the analytical biophysical model connecting the mGRE signal with biological tissue microstructure. \fb{Rev 2 Com 1}\textcolor{xiaojian}{LEARN-BIO is related to our recent method RoAR~\cite{Torop.etal2020} that trains CNNs to learn a mapping from Gaussian noise corrupted mGRE images to noise-free $R_2^\ast$ maps. However, unlike LEARN-BIO, RoAR does not account for motion during training,} \textcolor{yablonskiy}{which is the focus of the current work.}

Both of our approaches, LEARN-IMG and LEARN-BIO,  are trained on pairs of motion-corrupted and motion-free MR images without requiring the ground truth quantitative $R_2^\ast$ maps, which might be challenging to obtain in some settings. The advantage  of LEARN-IMG is that its training is decoupled from the quantitative mapping procedure, which means that the training does not need any prior knowledge of the biophysical model or pre-estimation of the F(t) functions. Instead, LEARN-IMG is used to produce motion-artifact-free mGRE images. Consequently, it has the flexibility to enable the successive estimation not just $R_2^\ast$ but also various mGRE-based quantitative maps (e.g. $R2t^\ast$~\cite{Ulrich.Yablonskiy2016, Zhao.etal2016}, cellular density~\cite{Wen.etal2018a}). The key advantage of LEARN-BIO is that by using  $F(t)$ during training, it learns to compensate for macroscopic magnetic field inhomogeneities to produce motion-artifact-free and $B0$-inhomogeneity-corrected $R_2^\ast$ maps. As a result, the trained LEARN-BIO can be directly applied to the motion-corrupted mGRE images without pre-computing $F(t)$ or using any additional fitting methods, resulting in a much faster computation of the quantitative maps. 

We train both of our CNN models on synthetic motion-corrupted data generated using our motion simulation pipeline. We show that both approaches are capable of removing motion artifacts on synthetic as well as experimental datasets and produce high quality in vivo quantitative maps. Quantitative and qualitative evaluations are conducted to demonstrate the robustness and effectiveness of the proposed methods. 

\section*{Methods}

\subsection*{The mGRE sequences and biophysical model}
\label{sec:biophysicalmodel}
In the $R_2^\ast$ approximation, the mGRE signal from a single voxel can be expressed as~\cite{Yablonskiy1998}:
	\begin{equation}
	\label{Eq:Model}
	S(t) = S_0 \cdot \exp(-R_2^\ast \cdot t - i \omega t) \cdot F(t),
	\end{equation}
	where $t$ denotes the gradient echo time, $S_0 = S(0)$ is the signal intensity at $t = 0$, and $\omega$ is a local frequency of the MRI signal. The complex  valued function $F(t)$ in Eq.~\eqref{Eq:Model} describes the effect of  \fbb{R1.1}\textcolor{xiaojian2}{intra- and inter-voxel} macroscopic magnetic field inhomogeneities on the mGRE signal. The failure to account for such inhomogeneities is known to bias and corrupt the recovered $R_2^\ast$ maps. In this paper we use the Voxel Spread Function (VSF) approach~\cite{Yablonskiy.etal2013} for calculating $F(t)$. \fb{Rev 1 Com 6}\textcolor{yablonskiy}{In the Voxel Spread Function (VSF) approach~\cite{Yablonskiy.etal2013}, effects of macroscopic magnetic field inhomogeneities (background gradients) on formation of mGRE signal are evaluated from the same complex mGRE dataset that is used for calculating tissue-specific parameters of a biophysical model. They are accounted for by including in the biophysical model the term F(t) that is calculated for each imaging voxel based on the values of mGRE signal phase and amplitude in this and surrounding voxels. The latter is important due to amplified signal leakage effects from the neighboring voxels (Fourier leakage) in the presence of magnetic field inhomogeneities~\cite{Bashir.Yablonskiy2006}.}  In a standard approach, the $R_2^\ast$ maps, $\omega$ maps, and $S_0$  are jointly estimated from 3D mGRE signals acquired at different echo times $t$ by fitting Eq.~\eqref{Eq:Model} with pre-calculated $F(t)$ on a voxel-by-voxel basis to experimental data by applying the non-linear least squares (NLLS) analysis. However, mGRE images are often affected by motion artifacts resulting from subject movement during MRI scan. In this paper we propose two learning-based approaches that can compute motion- and $B0$-inhomogeneity-corrected  $R_2^\ast$ maps from motion-corrupted mGRE data.


\subsection*{Motion artifacts simulation procedure }
\label{sec:motion_simulation}
Here we present our motion generation pipeline where various levels of motion artifacts can be introduced into motion-free MR images through manipulation of k-space data, which allows us to obtain pairs of motion-free and motion-corrupted images for training our CNNs. Specifically, denote the spatial motion-free mGRE images of $N$ echo times (e.g. $N =10$ in our data) at slice index $\ell$ as 
\begin{equation}
\label{Eq:sl}
\sbf_\ell = (\sbf_\ell^1, \dots, \sbf_\ell^N),
\end{equation}
where each component $\sbf_\ell^n  \in \mathbb{C} ^{y \times z}$ in $\sbf_\ell$ denotes a  2D complex image extracted from 3D  MR volume at slice $\ell$ for one of the echo times. Let $\ubf_\ell \in \mathbb{C} ^{k_y \times k_z \times N}$ denote the k-space maps of $\sbf_\ell$ such that 
\begin{figure}[t]
	\centerline{\includegraphics[width=\textwidth]{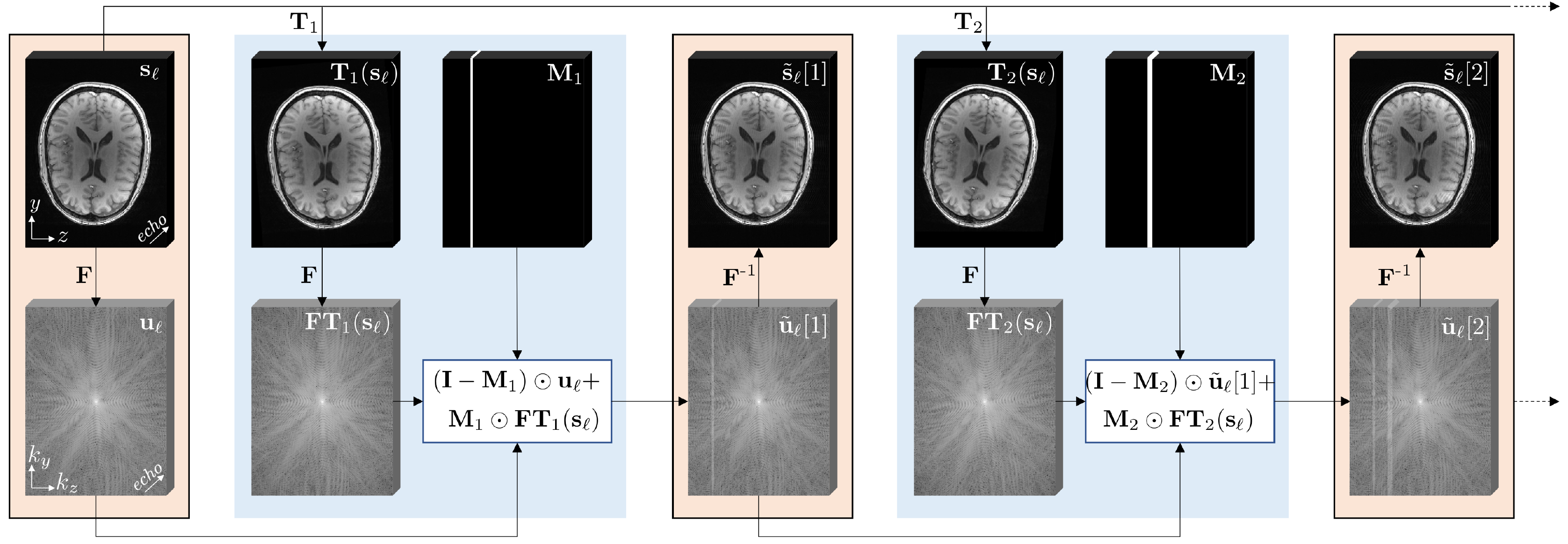}}
	\caption{Illustration of the motion corruption simulation pipeline for given example images $\sbf_\ell$. We assume all $N$ echoes at the given slice $\ell$ are corrupted by the same motion artifacts.  $\Tbf_1$ models the first shift and rotation when certain k-space lines at $\ubf_\ell$, indicated by $\Mbf_1$, are replaced by the k-space lines of the moved object,  generating the first $N$ echoes of motion-corrupted images $\sbftilde_{\ell}[1]$. Similarly, the second motion represented by  $\Tbf_2$ and $\Mbf_2$ further corrupts $\sbftilde_{\ell}[1] $ by generating  images $\sbftilde_{\ell}[2]$ with stronger artifacts. The process can be repeated a desired number of times.}
	\label{Fig:pipeline}
\end{figure}
\begin{align}
\label{Eq:yl}
\ubf_\ell = \Fbf \sbf_{\ell}  \quad \text{and} \quad
\sbf_\ell = \Fbf^{-1} \ubf_{\ell},
\end{align}
where $\Fbf $ and $\Fbf^{-1}$  represent the  Fourier  and inverse Fourier transforms. We model the motion artifacts in the MR images as the consequences of a series of physical motions, such as shifts or rotations, that result in 
perturbations of blocks of k-space lines during corresponding motions. We therefore replace certain k-space lines of the motion-free MR images with those of their motion-corrupted versions to synthesize motion artifacts. Specifically, in our data, for each point in $k_z$, we collect $N = 10$ echoes of k-space data along the $k_y$ direction in $\ubf_\ell$. Considering the fact that k-space scanning in the frequency-encoding direction ($k_y$) is much faster than the physical movement, we assume that all $10$-echo lines along $k_y$  suffer from the same motion effects (it takes about 50 ms to get 10 gradient echoes in our approach - see dataset section). We illustrate this motion generation method in a 2D slice-based manner which can be easily extended to 3D mGRE data by including the slice dimension. In particular, we represent the motion artifacts within a scan of slice $\ell$ as 
\begin{equation}
\label{Eq:Top}
\Tbf = (\Tbf_1, ..., \Tbf_J )
\end{equation}
where $\Tbf_j$ denotes the $j$th motion movement during the k-space data collection of $\sbf_\ell$ and $J$ is the total number of motions.  Let the binary map $\Mbf_j$ indicate the frequencies that are miscollected during the $j$th motion (with 1 in those k-space locations and 0 otherwise), simulating the time and duration of each sudden motion. Then, the final k-space maps after $J$ motions can be computed as 
\begin{equation}
\label{Eq:ybftilde}
\ubftilde_{\ell} [J] = (\Ibf - \sum_{j = 1}^{J} \Mbf_j) \odot \ubf_\ell+ \sum_{j = 1}^{J}  \Mbf_j \odot \Fbf \Tbf_j(\sbf_{\ell}), 
\end{equation}
where $\Ibf$ denotes a binary map with all ones and $\odot$ denotes the element-wise multiplication of two maps. As a result,  our synthetic  motion-corrupted images with $J$ motions can be computed as 
\begin{equation}
\label{Eq:sbftilde}
\sbftilde_{\ell} [J] = \Fbf^{-1}(\ubftilde_{\ell}[J])
\end{equation}
By changing the total  number of motions $J$ in addition to the location and duration of each physical motion indicated by $\Mbf_j$, one can control the  type and level of motion artifacts introduced to the motion-free images, and thus synthesize a variety of realistic motion-corrupted images. Notably, motion artifact generation can be conducted sequentially as the relationship between motion-corrupted images $\sbftilde_{\ell}[j-1]$ with the first $j-1$ motions and $\sbftilde_{\ell}[j]$ with the first $j$ motions in $\Tbf$ is 
\begin{equation}
\label{Eq:s_rela}
\sbftilde_{\ell}[j] = \Fbf^{-1} ((\Ibf - \Mbf_j) \odot \Fbf (\sbftilde_\ell[j-1] )+  \Mbf_j \odot \Fbf \Tbf_j(\sbf_{\ell})).
\end{equation}

This k-space lines-replacement-based motion artifacts generation pipeline is used to simulate the motion-corrupted data for training our CNNs. \fb{Rev 1 Com 1}\textcolor{xiaojian}{Specifically, we focus on the artifacts introduced by in-plane translational and 3D rotational movements where the subject is assumed to lie still during the examination with several swift translations or rotations of the head occurring during the process.} It is important to note that although the effect of global motion on the acquired k-space data is well established where translational motion induces a linear phase shift and rotational motion causes the same degree of rotation in the k-space data, our pipeline is actually more flexible as we can allow the k-space manipulation for more motion types (i.e., deformation and scaling) of different organs and we are not restricted to global motions. Figure~\ref{Fig:pipeline} illustrates an example where the motion artifacts are due to two consecutive rigid motions.

\subsubsection*{Method 1: LEARN-IMG}
\begin{figure}[t]
	\centerline{\includegraphics[width=.8\textwidth]{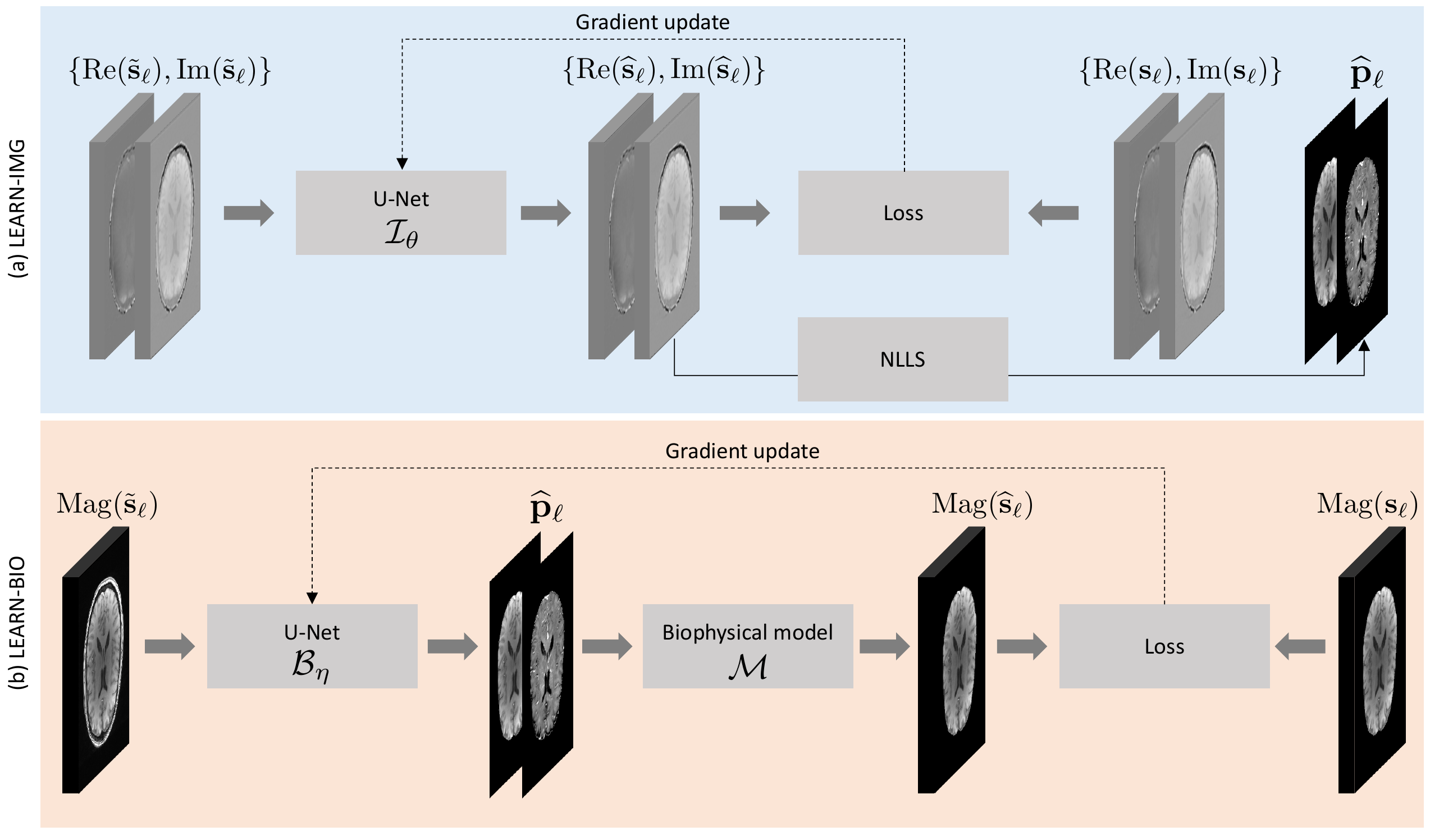}}
	\caption{Comparison of two approaches for training the motion-correction models. (a) In LEARN-IMG, the model $\Ical_\thetabm$ is optimized so that $\{\operatorname{Re}(\sbfhat_\ell), \operatorname{Im}(\sbfhat_\ell)\} = \Ical_\thetabm(\{\operatorname{Re}(\sbftilde_\ell), \operatorname{Im}(\sbftilde_\ell)\})$ is close to the corresponding ground-truth mGRE data $\{\operatorname{Re}(\sbf_\ell), \operatorname{Im}(\sbf_\ell)\}$. Standard NLLS approach is applied to the motion-corrected output  $\sbfhat = \{\operatorname{Re}(\sbfhat_\ell), \operatorname{Im}(\sbfhat_\ell)\} $ to estimate  unknown motion-free quantitative maps $\pbf_\ell = (\Sbf_{0}, \Rbf_{2}^\ast )$. (b) In the self-supervised approach LEARN-BIO,  the model $\Bcal_\etabm$ is trained to directly estimate motion-free quantitative maps $\pbf_\ell$ given measurement mGRE data $\operatorname{Mag}(\sbftilde_\ell)$ and the biophysical model $\Mcal$. The loss if formulated so that $\operatorname{Mag}(\sbfhat_\ell) = \operatorname{Mag}(\Mcal (\Bcal_\etabm (\operatorname{Mag}(\sbftilde_\ell)); \fbf_\ell))$ is close to corresponding ground-truth mGRE data $\operatorname{Mag}(\sbf_\ell)$.}
	\label{Fig:models}
\end{figure}

As defined in Eq.~\eqref{Eq:sl}, given the motion-free complex spatial mGRE images of $N$ echo times at slice $\ell$ as $\sbf_\ell$, we represent the corresponding absolute value of $S_0$ and true $R_2^\ast$ maps as
\begin{equation}
\label{Eq:pbf} 
\pbf_\ell = (\Sbf_{0}, \Rbf_{2}^\ast ).
\end{equation}
Let $\{\operatorname{Re}(\sbfhat_\ell), \operatorname{Im}(\sbfhat_\ell)\} = \Ical_\thetabm(\{\operatorname{Re}(\sbftilde_\ell), \operatorname{Im}(\sbftilde_\ell)\})$ denote our neural network LEARN-IMG \label{key} that computes an estimate $\sbfhat_\ell$ of the unknown motion-free $\sbf_\ell$ given the motion-corrupted mGRE signal $\sbftilde_\ell$.  The operators $\operatorname{Re}(\cdot)$ and  $\operatorname{Im }(\cdot)$ denote the real and imaginary parts of a complex number,  and vector $\thetabm$ denotes the trainable set of weights in the CNN.  The 3D convolutional structure of LEARN-IMG allows it to take both the complex statistical relationships between pixels and the echo times into account, and therefore enhances the motion correction performance  and robustness of the model. As illustrated in Figure~\ref{Fig:models}(a), the training of our network is carried out by minimizing the empirical loss over a training set consisting of $L$ slices $\{(\sbftilde_\ell, \sbf_\ell)\}_{\ell = 1, \dots, L}$, as follows
\begin{equation}
\label{Eq:cnnimg-loss}
\min_{\thetabm} \sum_{\ell = 1}^L \Lcal(\Ical_\thetabm(\{\operatorname{Re}(\sbftilde_\ell), \operatorname{Im}(\sbftilde_\ell)\}), \{\operatorname{Re}(\sbf_\ell), \operatorname{Im}(\sbf_\ell)\}),
\end{equation}
where $\Lcal$ measures the discrepancy between the vectorized estimates $\sbfhat_\ell$ generated by the LEARN-IMG and the ground-truth $\sbf_\ell$ on both imaginary and real channels. Common choices for $\Lcal$ include the $\ell_1$ and the $\ell_2$ distances. This minimization problem can be 
solved by using stochastic gradient-based optimization algorithms such as Adam~\cite{Bottou2012, Kingma.Ba2015}.

Our CNN architecture processes the 3D volumetric image of the whole brain by applying the model slice by slice.  Once the optimal set of parameters $\thetabm^\ast$ are learned from minimizing the optimization problem on the training dataset, which consists of many slices, the trained network $\Ical_{\thetabm^\ast}$ can be applied to unseen data to perform motion-correction tasks.
As illustrated in Figure~\ref{Fig:models}(a), the output of a LEARN-IMG network on the motion-corrupted images $\sbftilde_\ell$ is the  motion-corrected  complex mGRE data $\sbfhat_\ell$, decomposed with its real and imaginary components. The unknown motion-free quantitative maps $\pbf_\ell$ of input signals $\sbftilde_\ell$ can then be obtained by feeding the output signals $\sbfhat_\ell$  into the standard NLLS analysis, where both magnitude and phase images are needed to compute $F(t)$ values used during NLLS fitting. Notice this NLLS approach is only used for the quantitative maps computation during the test stage, not for the training of our CNN model.

\subsubsection*{Method 2: LEARN-BIO}
Let $\pbfhat_\ell = \Bcal_\etabm(\operatorname{Mag}(\sbftilde_\ell))$ denote our model LEARN-BIO, which computes the estimates $\pbfhat_\ell$ of the unknown true values of $\pbf_\ell$ given the magnitude value of the mGRE signal $\sbftilde_\ell$.  The operator $\operatorname{Mag}(\cdot)$  denotes the magnitude of the complex data and vector $\etabm$ denotes the trainable set of weights in the CNN. The network takes absolute data $\operatorname{Mag}(\sbftilde_\ell)$ as its $N$-channel input and produces $\pbfhat_\ell = (\widehat{\Sbf}_0, \widehat{\Rbf}_2^\ast)$ as its 2-channel output. The major difference between LEARN-BIO and the previously described LEARN-IMG is that the former directly learns to predict motion-free quantitative $R_2^\ast$ maps. LEARN-BIO is trained using a \emph{self-supervised learning} strategy, illustrated in Figure~\ref{Fig:models}(b), where the  model $\Bcal_\etabm$ is trained only using the mGRE data. \fb{Rev 1 Com 4 \\Rev 2 Com 3}\textcolor{xiaojian}{We adopt a widely-used definition of self-supervised learning where a model is trained using a \emph{pretext} (or \emph{auxiliary}) task, but tested on the actual \emph{desired} task~\cite{Yaman.etal2020a,Chen.etal2019a,Kim.Ye2020,Senouf.etal2019,Lu.etal2019, Gurbani.etal2019, Torop.etal2020}. In the context of our work, the \emph{desired} task is the estimation of $R_2^\ast$ maps, while the \emph{pretext} task is the generation of high-quality mGRE images.}  In LEARN-BIO, self-supervised learning is enabled by using the analytical biophysical model $\sbf_\ell = \Mcal(\pbf_\ell ; \fbf_\ell)$ in Eq.~\eqref{Eq:Model} to relate the mGRE images and the quantitative $R_2^\ast$ maps into a loss function and solving the corresponding optimization problem on a training dataset $\{(\sbftilde_\ell, \sbf_\ell)\}_{\ell = 1, \dots, L}$
\begin{equation} 
\label{Eq:cnnbio-loss}
\min_{\etabm} \sum_{\ell = 1}^L \Lcal\left(\operatorname{Mag}(\Mcal(\Bcal_\etabm(\operatorname{Mag}(\sbftilde_\ell)); \fbf_\ell)), \operatorname{Mag}(\sbf_\ell) \right),
\end{equation}
where $\fbf_\ell$ denotes matrices containing F-function values pre-calculated from the training mGRE  images $\sbf_\ell$ using the VSF method. Therefore, the training of LEARN-BIO is \emph{exclusively} reliant on the measurement data $\sbf_\ell$ instead of the ground-truth data $\pbf_\ell$, and is classified as a self-supervised method for using the supervision of measurements themselves through the signal model $\Mcal$ and the prior induced by the CNN to solve the model-fitting and motion-correction problem together.

As illustrated in Figure~\ref{Fig:models}(b), the output of our LEARN-BIO network for the $\ell$-th data element yields the quantitative maps $\pbfhat_\ell = \Bcal_\etabm(\operatorname{Mag}(\sbftilde_\ell))$,  which serve as an intermediate result for our optimization problem defined in Eq.~\eqref{Eq:cnnbio-loss}. The network is trained to find the best parameters $\etabm \ast$ such that its prediction $\{\pbfhat_\ell \}$ can well describe the measured mGRE signals and reduce the motion artifacts at the same time. \fb{Editor}\textcolor{xiaojian}{The detailed network structures of LEARN-BIO and LEARN-IMG are illustrated in the Supporting Information Figure S1.} As described below, the training of our CNNs is done by simulating the motion artifacts. 

\subsection*{Motion correction with LEARN-IMG and LEARN-BIO }
LEARN-IMG and LEARN-BIO are trained to remove motion artifacts from motion-corrupted mGRE data in order to produce high-quality $R_2^\ast$ maps. This can be achieved by solving the optimization problem defined in Eq.~\eqref{Eq:cnnimg-loss} and Eq.~\eqref{Eq:cnnbio-loss} using paired mGRE images $\{\sbftilde_\ell, \sbf_\ell\}$, where $\{\sbf_\ell\}$ are the images that are not contaminated by motion artifacts and $\{\sbftilde_\ell\}$ are the corresponding synthetic motion-corrupted images consisting of different levels of motion artifacts. We show below that our CNNs \emph{only} trained on synthetic data can achieve excellent performance on previously unseen experimental data. 

\subsection*{In vivo brain dataset}
For validating our method, we selected 20 different MRI scans with no visible motion artifacts (qualitatively inspected) from the previously published brain image data~\cite{Wen.etal2018a} as the motion-free source to generate the synthetic motion-corrupted  mGRE images. These brain image data are collected from 20 healthy volunteers (age range 26-76) using a Siemens 3T Trio MRI scanner and a 32-channel phased-array head coil. Studies were conducted with the approval of the local IRB of Washington University. All volunteers provided informed consent. The data was obtained using a 3D version of the mGRE sequence with $N = 10$ gradient echoes followed by a navigator echo~\cite{Wen.etal2015} used to reduce artifacts induced by physiological fluctuations during the scan. Sequence parameters were flip angle $FA = 30^\circ$, voxel size of $1 \times 1 \times 2$ mm$^3$, first echo time $t_1 = 4$ ms, echo spacing $\Delta t = 4$ ms (monopolar readout), repetition time TR $= 50$ ms, and the total imaging time for each acquisition was around 10 min.

In addition, experimental brain mGRE image data of 4 volunteers (ages:  32y, 67y, 71y, 78y) with clear visible motion artifacts were selected for evaluating the correction of real motions of our CNN models that trained on syntactic data.

\subsection*{Data generation pre-processing}
To obtain the paired motion-corrupted and motion-free images for training, brain mGRE images of 20 different MRI scans with no visible motion artifacts described above were selected to serve as the  the ``motion-free'' reference images for the training and quantitative evaluation of our CNNs. We split this data into 14 datasets (75\%) for training, 3 for validation  (15\%) and 3 for testing (15\%), and our aforementioned motion simulation procedure was then applied to these datasets slice by slice to generate motion-corrupted images.

To generate a range of realistic motion artifacts for our training dataset,  we select the total number of motions occurring during data acquisition as a random number in the range from 1 to 10. \fb{Rev 1 Com 1}\textcolor{yablonskiy}{For each motion, we simulated random in-plane shifts within the range of 0 to 15 voxels followed by a combination of three random rotations along each axis relative to the center of a 3D mGRE data volume, where each rotation is within the range of 0$^{\circ} $ to 15$^{\circ}$. While the simulation setting above yields excellent performance in our experimental data, it can be adjusted for different applications.} The time at which each motion occurred and the duration it lasted were randomly generated as well. In particular, all motions were assumed to occur randomly throughout the whole examination process, and each of them is assumed to last for a random duration from about 3 seconds to 30 seconds, which would be equivalent to disturbing about 1 to 10 k-space lines in a single 2D slice. All random numbers mentioned above were uniformly generated in the given range, introducing various levels of motion artifacts to our training and validation dataset. Those synthesized motion-corrupted data together with their motion-free origins were used for the training of our CNNs. 

To quantitatively evaluate the performance of our trained CNNs across different levels of motion artifacts, we synthesized three motion types using our test dataset with motion settings such that the corresponding artifacts introduced by each can well represent the different levels of corruption that appear in our experimental data. In particular, we manipulate 8\%, 16\% and 24\% of the k-space data for each slice in our test dataset, respectively. We name the motions generated through each of these settings as \emph{light}, \emph{moderate}  and \emph{heavy}  motion, based on the levels of artifacts  they introduced to our motion-free mGRE data. Those three settings were only applied to our test dataset to generate images at certain corruption levels for validating the robustness and capacity of our trained CNNs against different motion levels. Examples of these three motion type images are shown in Figure~\ref{Fig:motions}.

\begin{figure}[t]
	\caption{Illustration of synthetic motion-corrupted images. The background and skull voxels are masked out for better visualization. The left three columns show the magnitude of the $1$st of 10 echoes of mGRE images corrupted with \emph{light}, \emph{moderate} and \emph{heavy} motions, respectively. The relative error (RE) of each image is shown in its bottom left corner. The right three columns show the absolute differences between the motion-corrupted images and the motion-free images used to synthesize them.}
	\label{Fig:motions}
\end{figure}
Prior to feeding the data into our CNNs, the top slices, that include slice oversampling images and the intracranial space images,  and the bottom slices that include brain stem images or highly corrupted images due to the macroscopic field inhomogeneities, were discarded in each data volume. \textcolor{black}{Specifically, we only used the middle \emph{brain slices} corresponding to slice 25 through 55 of a 72-slice MRI data, 20 through 50 of a 60-slice MRI data, 30 through 60 of a 88-slice MRI data for all the experiments and numerical evaluation.} This resulted in 4340 images for training, 930 for validation, and 930 for testing in the simulation. Each dataset was also normalized to improve compatibility of our CNNs with different scanners, parameters, and intensity values following the strategy introduced in~\cite{Torop.etal2020}, where the signal intensity of each given data volume was divided by the mean of the signal intensity of its middle slice in the first echo. Consequently, the estimations  from different approaches on $S_0$ maps were scaled accordingly,  while $R_2^\ast$ maps were not affected.

\subsection*{Performance evaluation}
We trained our neural networks on a GeForce RTX 2080 GPU (NVIDIA Corporation, Santa Clara, CA, USA), and implemented in TensorFlow~\cite{Abadi.etal2016}, using the Adam optimizer to minimize the Euclidean distance. LEARN-BIO was trained for about 400 epochs (4 hours) and LEARN-IMG  for 200 epochs (24 hours) in order to achieve the best performance on the validation set. \fb{Editor}\textcolor{xiaojian}{Additional details on training are provided in the Supporting Information.} In the training of LEARN-BIO, the $F(t)$ function defined in Eq.~\eqref{Eq:Model} was used in the loss function to account for the macroscopic magnetic field inhomogeneities, which is essential in the estimation of $R_2^\ast$ maps free from $B0$ inhomogeneity artifacts. This $F(t)$  function was pre-computed using the VSF approach~\cite{Yablonskiy.etal2013} on the corresponding ground-truth images. Note that this $F(t)$ function was \emph{only} used during the training stage, and is \emph{not} required for testing purpose once the CNN is trained. This means our LEARN-BIO model is able to directly generate both motion- and inhomogeneity-corrected $R_2^\ast$ maps from the magnitude mGRE images. 

We used the traditional voxel-wise NLLS approach on the output of LEARN-IMG to produce corresponding motion-corrected $R_2^\ast$ maps.  NLLS is a standard iterative fitting method for computing  $R_2^\ast$ based on Eq.~\eqref{Eq:Model}. The $F(t)$ function computed using the VSF method on motion-corrected output of LEARN-IMG was also used before running NLLS in order to account for the effects of macroscopic magnetic field inhomogeneities. At each iteration, the regression is conducted by combining the data from different echo times $t$ with their $F(t)$ values voxel by voxel. Prior to the NLLS fitting procedures, a brain extraction tool, implemented in the Functional Magnetic Resonance Imaging of the Brain Library(FMRIB), was used to mask out both skull and background voxels in all MRI data~\cite{Jenkinson.etal2005}, where the signal model defined in Eq.~\eqref{Eq:Model} doesn't apply. NLLS, implemented in MatLab R2020a (MathWorks, Natick, MA),  was  run over only the set of unmasked voxels, optimizing for 400 iterations at each spatial point.  Similarly, we applied the same brain masks in the loss functions Eq.~\eqref{Eq:cnnimg-loss} and Eq.~\eqref{Eq:cnnbio-loss} for the training of our neural networks. Note for LEARN-BIO, those masks were \emph{only}  used during the training and were \emph{not}  needed for testing. All the visual results presented in this paper were also processed by these masks for better comparison. 

To demonstrate the performance of our proposed CNNs, the predicted $R_2^\ast$ results were compared with the ones computed from motion-corrupted mGRE data using the NLLS approach with a corresponding $F(t)$ function. In the synthetic scenario,  the $R_2^\ast$  maps computed from the motion-free mGRE images through NLLS can be thought of as a reference, which we only use at test time to quantitatively evaluate the $R_2^\ast$ results of different approaches. We use the \emph{relative error (RE)} metric and \emph{structural similarity index (SSIM)}~\cite{ZhouWang.etal2004}  as two means to quantitatively compare the estimated result $\xbfhat$  with its reference $\xbf^\ast$. We define RE as 
\begin{equation}
\text{RE} =  \frac{\left\|\xbf^\ast -  \xbfhat\right\|} {\left\|\xbf^\ast\right\|} \times 100 \%, 
\end{equation}
where $\xbfhat$ and $\xbf^\ast$ represent the vectorized image
estimation and its ground-truth reference respectively,  and $\|\cdot\|$ denotes the standard Euclidean norm. In the synthetic scenario,  to evaluate the performance of different approaches on motion-free $R_{2}^\ast$ estimation,  we regarded $R_{2}^\ast$ computed using NLLS on motion-free mGRE data as the reference. To evaluate the motion correction performance of LEARN-IMG on mGRE signals, we use the motion-free mGRE images as references. Both RE and SSIM are computed on the brain voxels indicated by the aforementioned brain mask for each slice. In the experimental scenario, where the motion-free references are not available, we applied the models trained on the synthetic data and provided a qualitative visual comparisons of the different approaches. 

\begin{figure}[t!]
	\centerline{\includegraphics[width=.7\textwidth]{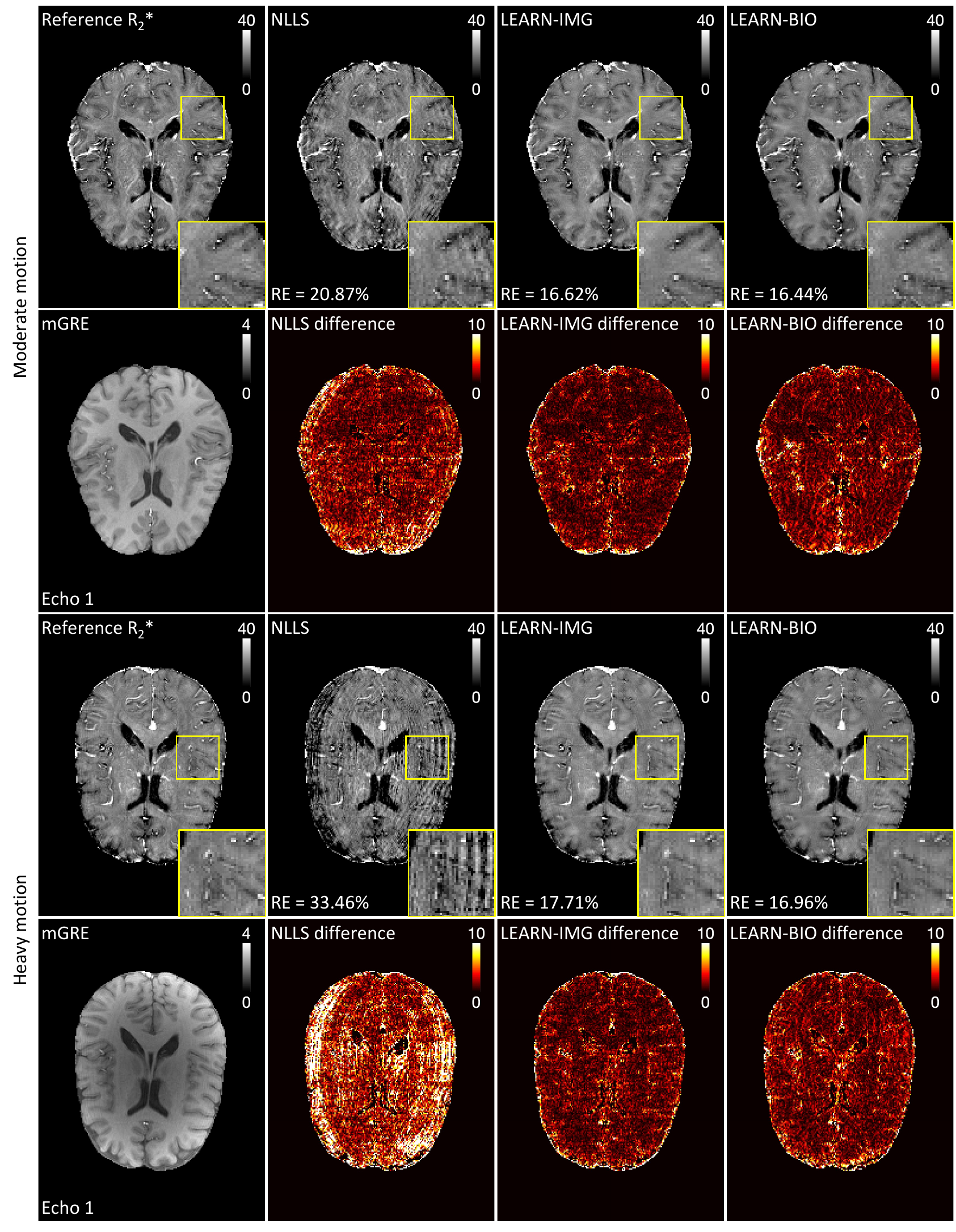}}
	\caption{Motion correction results from LEARN-IMG and LEARN-BIO on the synthetic data. Images represent two sets of results obtained from the synthetic datasets with different motion levels (\emph{moderate} in rows 1-2 and \emph{heavy}  in rows 3-4).  The Reference $R_2^\ast$ maps computed using NLLS on motion-free mGRE images and the magnitude of the $1$st of 10 echoes of motion-free mGRE images are shown in the first column.  Columns 2 to 4 show the $R_2^\ast$ maps and their absolute differences from the $R_2^\ast$ references for different methods. The relative error of each $R_2^\ast$ map is shown in the bottom left corner. Representative regions with $2\times$ zoom are shown on the bottom right of each $R_2^\ast$.}
	\label{Fig:res_sim}
\end{figure}

\section*{Results}
\label{sec:results}
\begin{figure}[H]
	\centerline{\includegraphics[width=.7\textwidth]{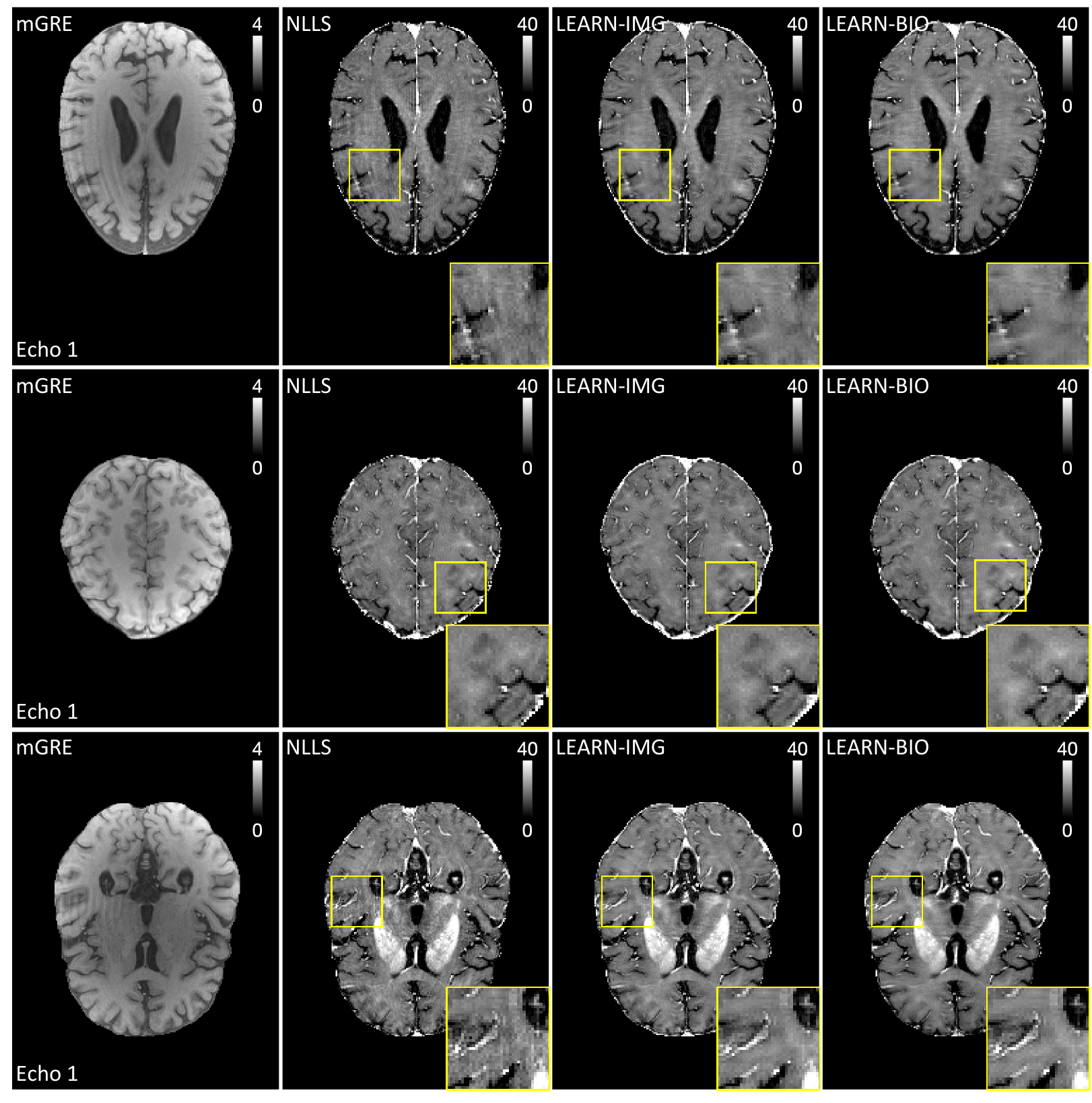}}
	\caption{\textcolor{xiaojian}{Motion correction results obtained using the experimental data from \emph{three} different subjects.} The magnitude of the $1$st of 10 echoes of motion-corrupted mGRE images of the method inputs are shown in the first column and $R_2^\ast$ maps of different methods are shown in column 2 to 4.  Representative regions with $2\times$ zoom are shown on the bottom right of each $R_2^\ast$.}
	\label{Fig:res_exp}
\end{figure}

Figure~\ref{Fig:res_sim} shows example $R_2^\ast$ maps calculated by NLLS, LEARN-IMG and LEARN-BIO for two simulated motion-corrupted slices in two different corruption scenarios (one is with moderate motions and the other is with heavy motions, as shown in Figure~\ref{Fig:motions}). Reference mGRE images without motion artifacts are shown for each example. It is clear that all the $R_2^\ast$ maps from NLLS contain strong motion artifacts, while the ones from LEARN-BIO and LEARN-IMG are of significantly higher quality with no obvious artifacts remaining. We also notice that while removing the artifacts, our approaches can also preserve important microstructure details of the $R_2^\ast$ maps, as can been seen from the exemplar zoomed regions in Figure~\ref{Fig:res_sim}. \fb{Rev 1 Com 1 }\textcolor{xiaojian}{Numerical RE results in each figure quantitatively corroborate that our methods provide better estimation of $R_2^\ast$ compared to NLLS,  with LEARN-BIO providing similar performance with LEARN-IMG in all examples. Note that the RE numbers in our results should be interpreted with care since the reference $R_2^\ast$ maps were computed using NLLS on motion-free in vivo data. Thus, despite the very similar RE performance compared to LEARN-IMG, LEARN-BIO usually achieves better motion-artifact removal as corroborated by visual evaluation (see zoomed details in Figure~\ref{Fig:res_sim} and Figure~\ref{Fig:res_exp}).} The difference maps ($|$estimated - reference$|$) that illustrate the absolute value of the deviation from their reference are also visualized for each result. It can be seen that difference maps of NLLS are much brighter compared to LEARN-IMG and LEARN-BIO, especially in the heavy motion scenario where NLLS estimations are contaminated by severe motion artifacts.  

\begin{figure}[t]
	\centerline{\includegraphics[width=0.8\textwidth]{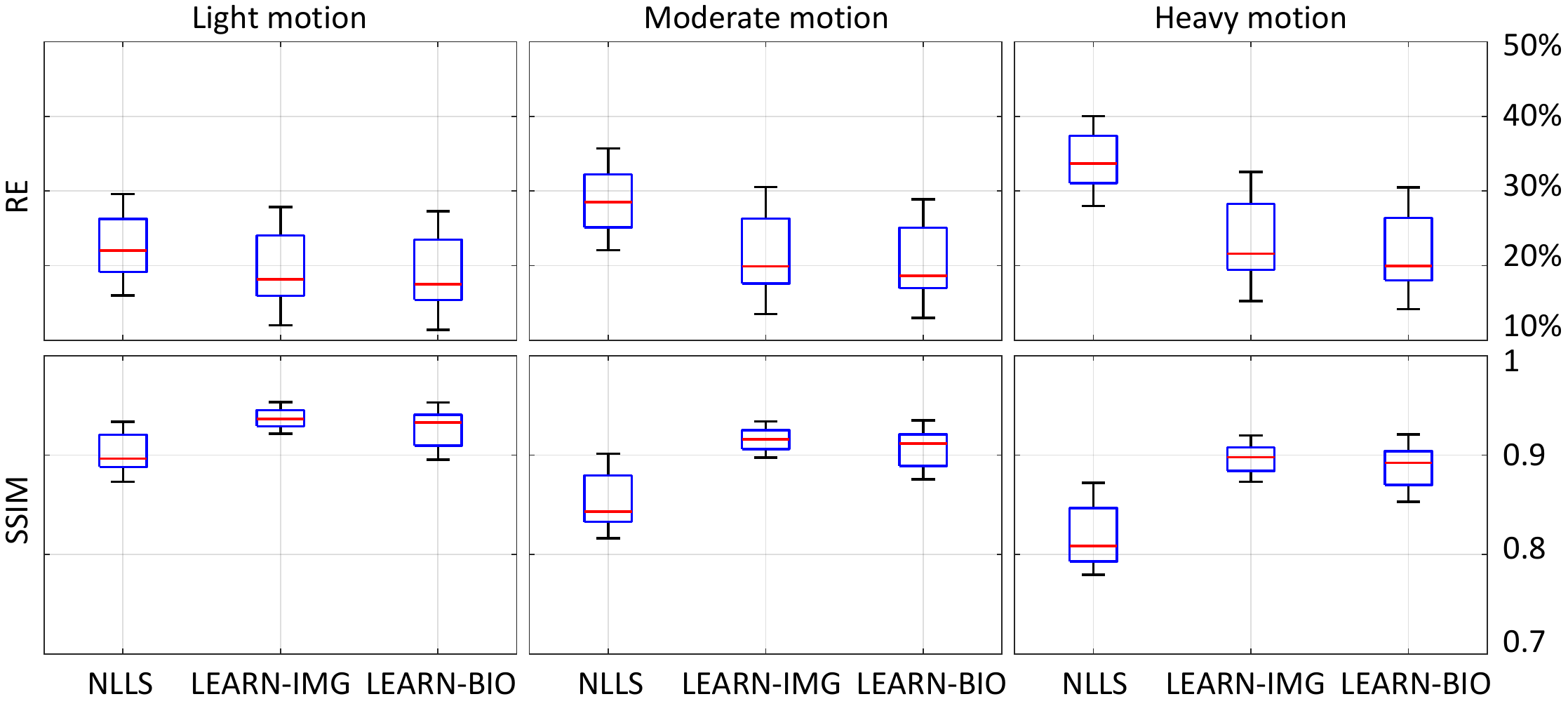}}
	\caption{The statistical analysis of NLLS, LEARN-IMG  and LEARN-BIO on the synthetic test data. Results visualize the performance of $R_2^\ast$ reconstruction from each approach in dealing with different levels of motion artifacts.}
	\label{Fig:statistics}
\end{figure}
Table~\ref{Tab:table_sim} and Figure~\ref{Fig:statistics} summarize the average RE  and SSIM over the whole test dataset on three motion corruption scenarios for all methods. We note that LEARN-IMG and LEARN-BIO share very similar performance across all corruption scenarios by providing much lower RE than NLLS. It is also worth noting that our proposed approaches are robust, showing about only $2\%$ gain of RE along with the increase of motion levels from light to moderate, and again from moderate to heavy. While the performance of the results from NLLS is dramatically affected by the motion levels,  showing about $6\%$ gain of RE along the increase of motion levels each time. As a results, the quality gap between our approaches and NLLS enlarges noticeably as inputs get more and more corrupted: from around 2\% at light to around 10\% at heavy. In addition to the $R_2^\ast$, the motion-corrupted mGRE inputs of the neural networks and the image predictions from LEARN-IMG are also evaluated and presented in the table. Here, the references for measuring the image predictions are the motion-free mGRE data and the numbers are computed on the magnitude of the complex data. The corresponding numerical results align with the conclusion over $R_2^\ast$, showing that LEARN-IMG is effective in removing motion artifacts in the spatial domain. 

\begin{figure}[H]
	\centerline{\includegraphics[width=\textwidth]{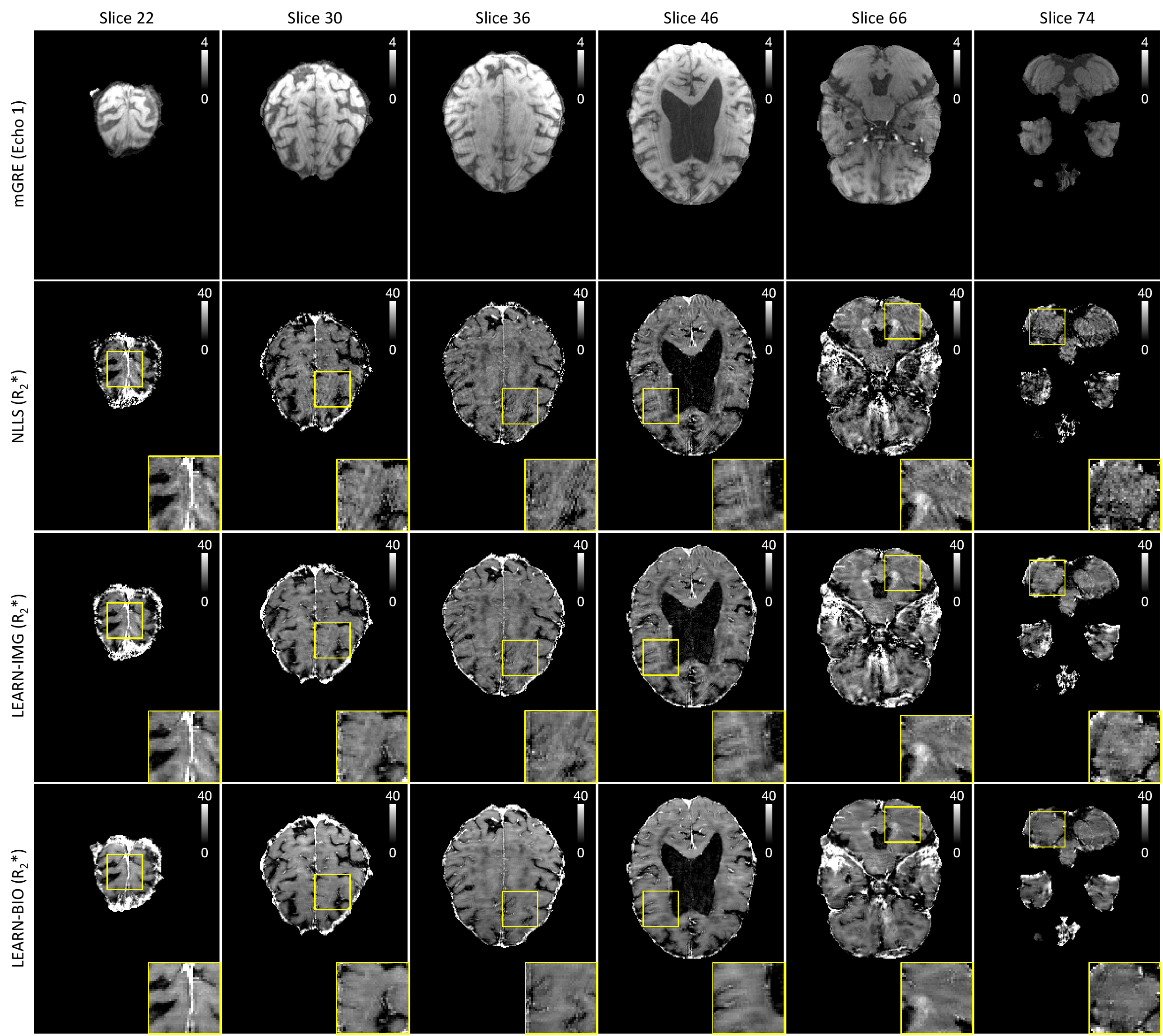}}
	\caption{\textcolor{xiaojian}{Motion correction results obtained using the experimental data from one subject across different brain slices. The representative top, middle,  and bottom slices of a 3D data volume are illustrated in columns 1-2, 3-4, and 5-6,  respectively. The magnitude of the $1$st of 10 echoes of motion-corrupted mGRE images are shown in row 1  and corresponding $R_2^\ast$ maps of different methods are shown in rows 2-4. Representative regions with $2\times$ zoom are shown on the bottom right.}}%
	\label{Fig:fullslice}
\end{figure}

\fb{Rev 1 Com 5}\textcolor{xiaojian}{Figure~\ref{Fig:res_exp} visualizes examples of the $R_2^\ast$  calculated by NLLS, LEARN-IMG, and LEARN-BIO for \emph{three} in vivo slices of \emph{different} subjects from the experimental data with real motion.} While the motion artifacts in this data might not follow our simulation model, we do observe similar results to our synthetic experiments. It can be seen that the $R_2^\ast$ maps produced by our methods are much better than the ones from NLLS, showing that our methodology is capable of handling the real motion artifacts while still keeping detailed structural information. \fb{Rev 1 Com 3}\textcolor{xiaojian}{While our networks are trained on the middle slices, they can perform motion correction across the whole brain volume (including the very top and bottom slices). As shown in Figure~\ref{Fig:fullslice}, our networks that were trained on middle slices can remove the motion artifacts across different brain regions, including the top (columns 1 and 2), middle (columns 3 and 4),  and bottom (columns 5 and 6) slices in the experimental data with real motion. Our networks outperform the baseline method NLLS which suffers from high motion artifacts in the estimated $R_2^\ast$ maps across the whole brain.} \fbb{R1.3}\textcolor{xiaojian2}{We hypothesize that the ability of our networks to generalize across different brain regions is due to their ability to capture similarity in the motion artifacts.} 

\fb{Rev 1 Com 5}

\fb{Rev 1 Com 3}

\section*{Discussion and Conclusions}
\label{sec:conclusion}
In this manuscript we proposed two convolutional neural network approaches, namely LEARN-IMG and LEARN-BIO,  for the robust motion correction of $R_2^\ast$  maps from multi-Gradient-Recalled Echo MRI data. LEARN-IMG is based on a supervised deep neural network that conducts the motion correction on complex mGRE images, followed by the standard NLLS fitting approach. It decomposes the motion correction and quantitative maps estimation into two procedures, bringing more flexibility to various potential applications of quantitative mapping. In this approach, the training of CNN simply relies on the paired MR images. In testing, it only takes 30 seconds for CNN to process the full brain data (using a GeForce GTX 1080 Ti GPU), while NLLS fitting together with the computation $F(t)$ functions take about 120 minutes on a modern PC (using 8 cores).  On the other hand, LEARN-BIO is based on a self-supervised  deep neural network that uses a biophysical model connecting mGRE MRI signal with underlying biological tissue microstructure. It integrates the motion correction and quantitative mapping procedures together within one single CNN. During the training (but not application), both paired MR images and the $F(t)$ functions as well as the knowledge of the biophysical model are needed. This allows our CNN to recognize the contribution of macroscopic magnetic field inhomogeneities to the mGRE signal only from the magnitude data. Therefore, at testing time, the information of $F(t)$ functions is not required, which saves a great amount of time compared with the LEARN-IMG approach. Specifically, it takes only 3 seconds for LEARN-BIO to predict the motion- and $B0$-corrected $R_2^\ast$ maps for the full brain data. 

Both of our methods, LEARN-IMG and LEARN-BIO, show great performance in producing motion- and $B0$-corrected $R_2^\ast$ maps that are of the same quality as NLLS-based voxel-by-voxel analysis from motion-free mGRE data. As shown in Figure~\ref{Fig:res_sim}  and Table~\ref{Tab:table_sim},  LEARN-BIO and LEARN-IMG consistently gives the best performance over the synthetic data in our different corruption scenarios. Figure~\ref{Fig:res_exp} and Figure~\ref{Fig:fullslice} further elaborates the capability of our CNN models on experimental data, showing a practical application of our approaches on removing real-world motion artifacts and keeping feature details. Our approaches constantly outperform NLLS both qualitatively and quantitatively, thanks to the power of our deep neural networks. 

\fb{Rev 1 Com 7}\textcolor{xiaojian}{While both of our proposed methods enable the estimation of high quality motion-corrected  $R_2^\ast$ maps, they have distinct trade-offs that can lead to different applicability in practical scenarios. LEARN-BIO is suitable for applications where one directly requires motion-correction quantitative $R_2^\ast$ maps with low computational time. \textcolor{yablonskiy}{In this paper, we used LEARN-BIO with a specific biophysical model in Eq.~\eqref{Eq:Model} to produce $R_2^\ast$ maps, but the same approach can potentially be applied to any biophysical model.} LEARN-IMG, on the other hand, \textcolor{yablonskiy}{provides motion-corrected mGRE images that can be used in a combination with any previously developed model that did not account for motion in the data.} \fb{Rev 2 Com 4}\textcolor{yablonskiy}{For example, motion-corrected mGRE images can be used for generating quantitative tissue-cellular-specific ($R2t^\ast$) and Blood-Oxygen-Level-specific ($R2'$) maps~\cite{Yablonskiy1998, Ulrich.Yablonskiy2016}, which have proved useful in studying healthy aging~\cite{Zhao.etal2016}, brain neuronal content~\cite{Wen.etal2018a} and its relationship to brain functional connectivity~\cite{Kahali.etal2021}, Alzheimer Disease~\cite{Zhao.etal2017a, Kothapalli.etal2021}, Multiple Sclerosis~\cite{Xiang.etal2019, Xiang.etal2020, Xiang2021.08.18.21262247}, and psychiatric disease~\cite{Mamah.etal2015}.} mGRE sequences are also used to study brain tissue multi-compartment structure~\cite{Du.etal2007, Sati.etal2013, Alonso-Ortiz.etal2018}, mapping of cerebral metabolic rate of oxygen by combining quantitative susceptibility mapping and quantitative blood oxygenation level-dependent imaging~\cite{Cho.etal2018}, etc. Therefore, LEARN-IMG can be generalized to various applications and different quantitative map computation. Nevertheless, this generalization capability requires applying additional quantitative-map-oriented fittings to the motion-corrected mGRE outputs of the network, which consequently increases the computation time compared to LEARN-BIO. Therefore, LEARN-IMG fits applications where different quantitative maps are required under less restrictive time constrains.}

\fb{Rev 1 Com 2}\textcolor{xiaojian}{In this work, we have accounted for the macroscopic magnetic field inhomogeneity effects by including in the biophysical model the F-function term (Eq.~\eqref{Eq:Model}) computed using the the Voxel Spread Function approach~\cite{Yablonskiy.etal2013} }\textcolor{yablonskiy}{that accounts for the presence of magnetic field inhomogeneities in the imaging object.} Since magnetic field inhomogeneities are induced mostly by the air cavities (e.g. sinuses) inside the head, they change with head motion, thus affecting MRI data. This effect is not accounted for in our current motion-correction implementation. \fbb{R1.5}\textcolor{xiaojian2}{While detail analysis of this effect is beyond the scope of this paper, we can estimate a potential error resulting from this effect by recalling that only a projection ($\Delta B$) of the magnetic susceptibility-induced inhomogeneous magnetic field ($\Delta\Bbf$) on the main magnetic field $B0$ actually affects MRI signal. Hence, if the object in the MRI scanner is rotated by an angle $\theta$, the $\Delta B$ would not change by more than $\cos^2\theta$. E.g. for $\theta$ about $10^\circ$, this change is only about 3$\%$ and can potentially cause about 3$\%$  error in $R_2^\ast$ estimation in the brain regions strongly affected by the magnetic field inhomogeneities (see Figure A2 in \cite{Kahali.etal2021}). However, in a typical MRI session, 3D mGRE data are acquired over about 6-minute scan while motion usually does not last for more than a few seconds, affecting only small portion of 3D dataset, and consequently resulting in $R_2^\ast$ estimation error significantly smaller than 3$\%$.}

\fbb{R1.2}\textcolor{xiaojian2}{It worth mentioning that while 3D motion is considered in the simulation, our network architectures takes as its input a sequence of 2D images at different echo times. The key benefit of using our architectures, instead of more complex ones that consider sequences of 3D images, is the lower computational and memory complexity. However, it conceivable that by using more complex architectures one can better capture the 3D motion and thus improve the final performance. \fbb{R1.4} Similar to all image restoration methods, our method exhibits a common trade-off between artifact-removal and smoothing on some images. In future work, it might be worth exploring the potential of obtaining sharper images by replacing the Euclidean loss used in our training with other common functions (e.g. $\ell_1$ loss).}

\textbf{In Conclusion,} we introduced LEARN-IMG and LEARN-BIO as two fast and robust learning-based methods that can utilize  motion-corrupted mGRE data to produce high quality $R_2^\ast$ maps which are free from
$B0$-inhomogeneity and motion artifacts. This validates the representation power of our convolutional neural networks in quantitative map estimation as well as motion correction. The good motion-reduction performance on experimental data demonstrates a potential clinical usage of our trained models.

\section*{Appendix}
\label{sec:appendix}

\textbf{Acknowledgment:} The authors are grateful to Max Torop, Alexander Sukstansky, for helpful discussion, and Kyle Singer for his aid in proofreading the paper.



\clearpage
\section*{Tables}

\begin{table}[H]
	\centering
	\caption{Average RE and SSIM values for the $R_2^\ast$ estimation on the synthetic test data. Results here summarize the performance of each approach at different levels of motion artifacts. We additionally provide the RE and SSIM values for the motion-corrupted  mGRE input data and the output of LEARN-IMG averaged on 10 echoes.}
	\vspace{5pt}
	\small
	\label{Tab:table_sim}
	\color{xiaojian}{
		\begin{tabular*}{14.9cm}{M{55pt}M{75pt}	 cC{27pt}C{27pt} cC{27pt}C{27pt}	cC{27pt}C{27pt}}
			\toprule
			\multirow{2}{*}{\textbf{}}& \multirow{2}{*}{\textbf{Method} }
			&&\multicolumn{2}{c}{\textbf{Light}}
			&&\multicolumn{2}{c}{\textbf{Moderate}}
			&&\multicolumn{2}{c}{\textbf{Heavy}} \\
			\cmidrule{4-5}\cmidrule{7-8}\cmidrule{10-11} 
			&
			&&\textbf{RE}&\textbf{SSIM}
			&&\textbf{RE}&\textbf{SSIM}
			&&\textbf{RE}&\textbf{SSIM}\\
			\cmidrule{1-11}
			
			\multirow{2}{*}{\textbf{mGRE}}
			& \textbf{Input}        		&&5.02\%&0.96		&&7.31\%&0.93		&&9.19\%&0.90		\\ 
			& \textbf{LEARN-IMG}         &&3.86\%&0.98		&&4.91\%&0.96		&&5.82\%&0.95		\\ 
			\noalign{\vskip 1pt}\cdashline{1-11}\noalign{\vskip 3pt}
			
			\multirow{3}{*}{\textbf{$R_2^\ast$}}
			& \textbf{NLLS}         		&&22.52\%&0.90	&&28.53\%&0.86	&&33.55\%&0.82	\\
			& \textbf{LEARN-IMG}         &&20.02\%&0.94	&&22.04\%&0.91	&&23.92\%&0.90		 \\
			& \textbf{LEARN-BIO}         &&20.28\%&0.93 	&&21.76\%&0.90	&&23.08\%&0.89 		 \\
			\bottomrule
		\end{tabular*}
	}
\end{table}
\fb{Rev 1 Com 1}


\listoffigures

\cftpagenumbersoff{figure}

\clearpage
\newpage

	\renewcommand{\suppfigurename}{Supporting Information Figure}
	\renewcommand{\thesuppfigure}{S\arabic{suppfigure}}
	
	\begin{figure*}[t]
		\begin{center}
			{\huge Supporting Information for ``Learning-based Motion Artisfact Removal Networks (LEARN) for Quantitative $R_2^\ast$ Mapping''}
		\end{center}
	\end{figure*}
	\newpage
	\maketitle
	\makeatletter
	
	
	
	\maketitle
	\subsection*{Architecture of CNNs}
	\begin{suppfigure}[t]
		\centerline{\includegraphics[width=\textwidth]{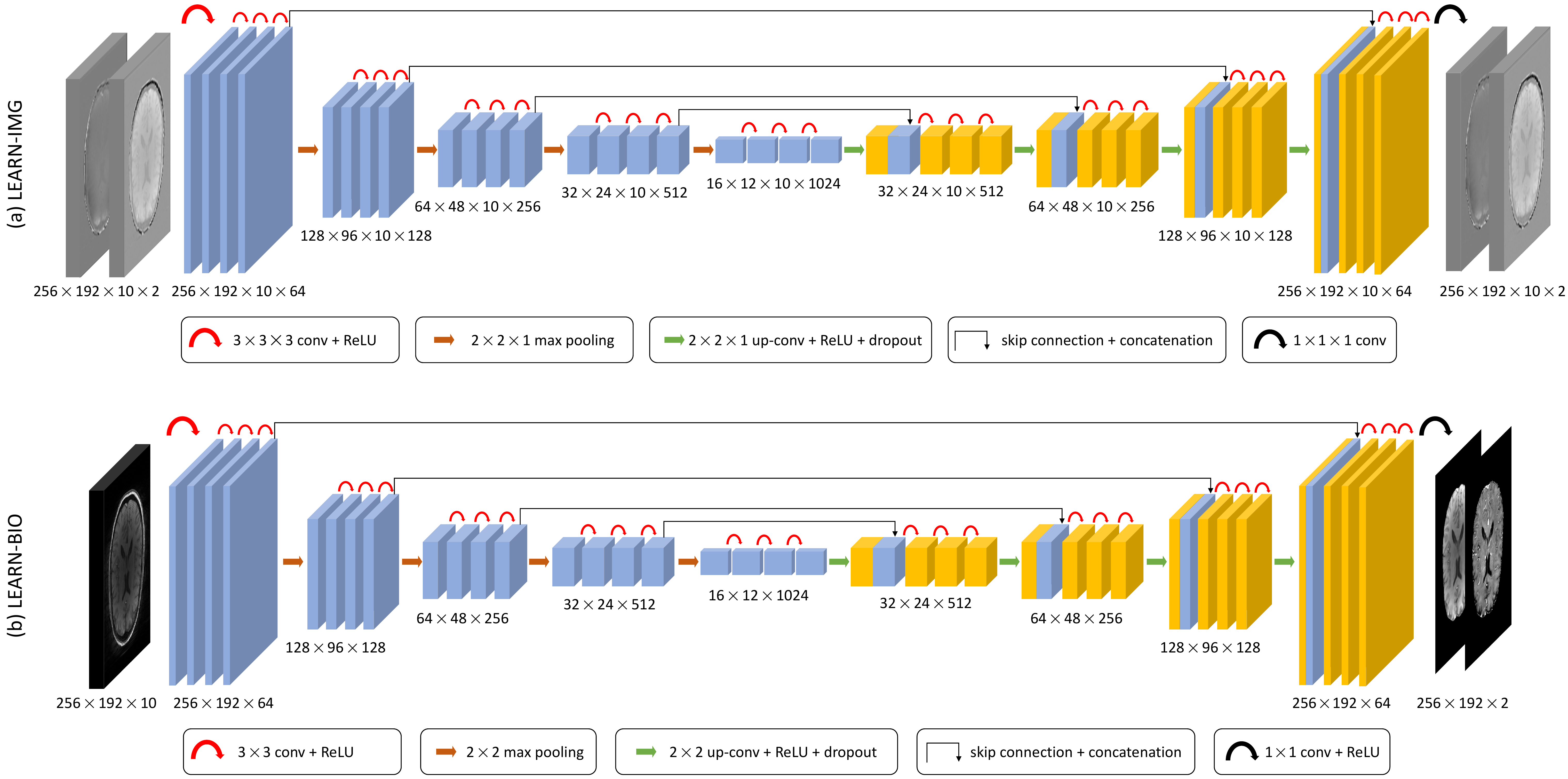}}
		\caption{CNN models for LEARN-IMG and LEARN-BIO. (a) In LEARN-IMG,  we split motion-corrupted complex mGRE $\sbftilde_\ell = (\sbftilde_\ell^1, \dots, \sbftilde_\ell^{10})$ into 2-channel data  as input $\{\operatorname{Re}(\sbftilde_\ell), \operatorname{Im}(\sbftilde_\ell)\}$ for neural network $\Ical_\thetabm$ to reconstruct the motion-corrected mGRE images $\{\operatorname{Re}(\sbfhat_\ell), \operatorname{Im}(\sbfhat_\ell)\}$ of the same size.   (b) In LEARN-BIO, neural network $\Bcal_\etabm$ takes 10-channel magnitude-only mGRE data $\operatorname{Mag}(\sbftilde_\ell) = (\operatorname{Mag}(\sbftilde_\ell^1), \dots, \operatorname{Mag}(\sbftilde_\ell^{10}))$  as input to reconstruct the 2-channel quantitative output $\pbfhat_\ell = (\Sbfhat_{0}, \Rbfhat_{2}^\ast)$.  Both models process data from individual spatial slices extracted from a 3D MRI data and the 3D image of the whole brain can be obtained by concatenating the outputs of the CNN applied slice-by-slice.}
		\label{Fig:cnns}
	\end{suppfigure}
	
	Supporting Information Figure~\ref{Fig:cnns}  presents the details of our LEARN-IMG and LEARN-BIO models, both of which are based on modifications to the popular  U-Net architecture~\cite{Ronneberger.etal2015, Cicek.etal2016}. The U-shaped architecture results from a combination of a contracting path and an expanding path, where the contracting path relies on a repeated usage of convolutions, each followed by a rectified linear unit (ReLU) and a max pooling operation to encode the spatial information, and the expanding path uses sequence of up-convolutions and concatenations with high-resolution features from the contracting path to increase the resolution of the output. The spatial information is reduced while feature information is increased during the contracting path, making the effective size of its filters in the middle layers larger than that of the early and late layers~\cite{Jin.etal2017a}. Such multi-scale structure leads to a large receptive field of the CNN that has been shown to be effective for removing globally spread imaging artifacts typical in MRI~\cite{Han.etal2017, Krizhevsky.etal2012}.
	
	\subsection*{Training of CNNs}
	We trained both of our networks on the synthetic motion-corrupted data with the following fine-tuning strategy. We first trained our networks, both LEARN-IMG and LEARN-BIO, on the 2D motion-corrupted data that synthesized based on the in-plane translational and rotational movements to achieve stable performance on our validation dataset. Such training process, which we refer as pre-training, took about 180 epochs for LEARN-IMG and  390 epochs for LEARN-BIO. We further fine-tuned both of our pre-trained network instances that achieved the best performance on our validation dataset with the 3D motion-corrupted data until stable convergence observed. The final network instances selected were based on the best performance on our validation dataset after fine-tuning step.  We noticed such a pre-training-and-fine-tuning strategy benefits the training stabilization and motion-correction performance in our experiments.

	\listofsuppfigures
\end{document}